\documentclass[aps,pra,twocolumn,groupedaddress,nofootinbib,longbibliography]{revtex4-1}
\usepackage{amsmath}
\usepackage{amssymb}
\usepackage{graphicx}
\usepackage{mathrsfs}
\usepackage{ntheorem}
\usepackage{mathtools}
\usepackage{enumerate}
\usepackage{enumitem}
\usepackage[T1]{fontenc}
\usepackage{physics}
\usepackage{subfig}
\usepackage{overpic}
\usepackage{epstopdf}
\usepackage{bm}
\usepackage{color,soul}
\usepackage[bookmarks=false]{hyperref}
\usepackage{xcolor}
\hypersetup{colorlinks=true,citecolor=blue,
linkcolor=blue,urlcolor=blue,pdfstartview=FitH,
bookmarksopen=true}

\begin{document}
\title{Maximal Steered Coherence in Accelerating Unruh-DeWitt Detectors}
\author{Hong-Wei Li$^{a,1}$, Yi-Hao Fan$^a$, Shu-Ting Shen$^a$, Xiao-Jing Yan$^b$, Xi-Yun Li$^b$, Wei Zhong$^c$, Yu-Bo Sheng$^{a,c}$}
\author{Lan Zhou$^b$}
\email{zhoul@njupt.edu.cn}
\author{Ming-Ming Du$^{a,1}$}
\email{mingmingdu@njupt.edu.cn}
\affiliation{$a.$ College of Electronic and Optical Engineering and College of Flexible Electronics (Future Technology), Nanjing
University of Posts and Telecommunications, Nanjing, 210023, China\\
$b.$ School of Science, Nanjing University of Posts and Telecommunications, Nanjing,
210023, China\\
$c.$ Institute of Quantum Information and Technology, Nanjing University of Posts and Telecommunications, Nanjing, 210003, China}
\footnote{the two authors contribute equally to this work}
\date{\today}
\begin{abstract}
Quantum coherence, a fundamental aspect of quantum mechanics, plays a crucial role in various quantum information tasks. However, preserving coherence under extreme conditions, such as relativistic acceleration, poses significant challenges. In this paper, we investigate the influence of Unruh temperature and energy levels on the evolution of maximal steered coherence (MSC) for different initial states. Our results reveal that MSC is strongly dependent on Unruh temperature, exhibiting behaviors ranging from monotonic decline to non-monotonic recovery, depending on the initial state parameter $\Delta_0$. Notably, when $\Delta_0=1$, MSC is generated as Unruh temperature increases. Additionally, we observe that higher energy levels help preserve or enhance MSC in the presence of Unruh effects. These findings offer valuable insights into the intricate relationship between relativistic effects and quantum coherence, with potential applications in developing robust quantum technologies for non-inertial environments.

\end{abstract}

\maketitle

\section{Introduction}
Quantum coherence, a fundamental feature of quantum mechanics, captures the ability of quantum systems to exhibit interference through the superposition of states \cite{Streltsov2017}. Beyond foundational physics, coherence plays a pivotal role in diverse areas, including quantum information science \cite{Streltsov2017,Yu2016,Yu2016a,Zhang2018,Du2022,Zhang2024}, thermodynamics \cite{Lostaglio2015, Cwiklinski2015, Lostaglio2015a, Narasimhachar2015}, and even biological processes \cite{Lloyd2011, Huelga2013}. However, preserving quantum coherence is a significant challenge due to interactions with the environment that often lead to decoherence. Studying quantum coherence under extreme conditions, such as relativistic acceleration, is particularly important for developing robust quantum technologies capable of operating in non-inertial. The Unruh effect \cite{Fulling1973, Davies1975, Unruh1976} indicates that an accelerating observer perceives the vacuum as a thermal bath, introducing additional decoherence. Understanding the influence of this effect on quantum coherence is crucial for advancing both fundamental insights and practical applications.

The influence of relativistic motion on quantum coherence has been extensively explored in recent studies. Early work in this directions \cite{Wang2016a} examined a system of two Unruh-DeWitt detectors, with one detector undergoing acceleration. The results indicated that quantum coherence undergoes irreversible degradation due to relativistic motion \cite{Wang2016a}. This important discovery subsequently inspired numerous investigations into the behavior of quantum coherence in various relativistic quantum systems \cite{Huang2018, Feng2021, Bhuvaneswari2022, Xiao2022, Harikrishnan2022, Du2024, Du2024a}. Furthermore, recent advancements have expanded the scope of these studies to include multipartite coherence in relativistic quantum systems \cite{Wu2019, Wu2021, Harikrishnan2022}, highlighting the broader implications of relativistic effects on more complex quantum states.

Despite considerable progress in understanding quantum coherence under relativistic conditions, the role of maximal steered coherence (MSC) in such environments remains largely unexplored. MSC, which represents the ability of one subsystem to control or influence the coherence of another \cite{Hu2016}, is a critical resource in quantum information processing \cite{Du2024a, Maleki2020, Xu2021}, particularly in the remote generation of quantum coherence. In non-inertial or accelerating frames, where decoherence effects due to the Unruh effect become pronounced, understanding the behavior of MSC is essential for advancing the robustness and reliability of quantum technologies.

In this study, we explore the dynamics of two accelerating Unruh-DeWitt detectors within a $3+1$-dimensional Minkowski spacetime \cite{Benatti2004}, modeling them as an open quantum system. By employing the Kossakowski-Lindblad master equation, we analyze the influence of critical parameters such as initial state $\Delta_0$, energy level $\omega$, and Unruh temperature $T$ on the evolution of MSC. Our results demonstrate that MSC is highly sensitive to Unruh temperature $T$, with behavior ranging from monotonic decay to non-monotonic revival depending on the initial state parameter $\Delta_0$. Furthermore, we find that higher energy levels $\omega$ can effectively preserve MSC in the presence of Unruh effects. These findings suggest that the Unruh effect exerts a complex influence on quantum coherence in open systems, with the potential to either diminish or augment MSC depending on the system's initial state parameters.

To provide a self-contained exposition, we organize the rest of this paper as follows. In Sec.
\Ref{sec2}, we provide a brief introduction to MSC. In Sec.
\Ref{sec3}, we describe the Unruh-DeWitt detectors model. In Sec.
\Ref{sec4}, we investigate the behavior of maximal steered coherence in Unruh-DeWitt detectors. Finally, we conclude this work in Sec.
\Ref{sec5}.

\section{The maximal steered coherence}\label{sec2}
Consider a bipartite quantum state $\rho$ shared between Alice and Bob. When Alice performs a measurement using a Positive Operator-Valued Measure (POVM) element $M$, Bob's subsystem is steered into a new state $\rho_{M_B}$. The quantum coherence of this steered state, $C(\rho_{M_B})$, is quantified by the sum of the absolute values of its off-diagonal elements in the eigenbasis $\{|\xi_i\rangle\}$ of Bob's reduced state $\rho_B$:

\begin{equation}
C(\rho_{M_B}, \{|\xi_i\rangle\}) = \frac{1}{p_M} \sum_{i \neq j} |\langle \xi_i|\text{tr}_A(M \otimes I \rho)|\xi_j\rangle|,
\end{equation}
where $p_{M_B} = \text{tr}(M \otimes I \rho)$ is the probability of obtaining the outcome associated with $M$. The MSC is defined as the maximum value of $C(\rho_{M_B})$ that can be achieved through all possible POVMs that Alice could perform \cite{Hu2016}:

\begin{equation}
C_{max}(\rho) := \max_{M \in \text{POVM}} \left[ \frac{1}{p_M} \sum_{i \neq j} |\langle \xi_i|\text{tr}_A(M \otimes I \rho)|\xi_j\rangle| \right].
\end{equation}

If $\rho_{M_B}$ is degenerate,  $\{|\xi_i\rangle\}$ is not uniquely defined; however, MSC is defined over all possible POVM operators and taking infimum over all possiblere ference basis as
\begin{align}
&C_{max}(\rho)\\\nonumber
&:= \inf_{\{|\xi_i\rangle\}} \left\{\max_{M \in \text{POVM}} \left[ \frac{1}{p_M} \sum_{i \neq j} |\langle \xi_i|\text{tr}_A(M \otimes I \rho)|\xi_j\rangle| \right]\right\}.
\end{align}
The MSC serves as a measure of the extent to which steering can influence and enhance quantum coherence.

\section{Unruh-deWitt detectors model}\label{sec3}
To elucidate the dynamics of two accelerating Unruh-DeWitt (UDW) detectors within $3+1$-dimensional Minkowski spacetime \cite{Benatti2004,Bhuvaneswari2022,Elghaayda2023}, this study models the detectors as an open quantum system, characterized by non-unitary evolution of the density matrix attributable to environmental decoherence and dissipation.

Each detector is modeled as a two-level atom, with the total Hamiltonian of the system expressed as follows:
\begin{equation}
H = \frac{\omega}{2} \Sigma_3 + H_\Phi + \mu H_I,
\end{equation}
where $\omega$ denotes the energy level spacing of the atom and $\mu$ is the coupling parameter. Here,
$\Sigma_i$ represents a symmetrized bipartite operator, defined as $\Sigma_i \equiv \sigma_i^{(A)} \otimes I^{(B)} + I^{(A)} \otimes \sigma_i^{(B)}$, with $\sigma_i^{(\alpha)}$ being the Pauli matrices, and $\alpha = A, B$ labels the different atoms.  $H_\Phi$ is the Hamiltonian of the free massless scalar field $\Phi(t, x)$, which satisfies the standard Klein-Gordon equation. The interaction Hamiltonian can be written in dipole form \cite{Hu2013}:
   \begin{equation}
    H_I = (\sigma_2^{(A)} \otimes 1^{(B)}) \Phi(t, x_1) + (1^{(A)} \otimes \sigma_2^{(B)}) \Phi(t, x_2)
   \end{equation}

Assuming weak coupling, characterized by the parameter $\mu \ll 1$, between the accelerated detectors and their environment, the initial state of the total system is represented as:
\begin{equation}
\rho_{\text{tot}}(0) = \rho_{AB}(0) \otimes |0\rangle \langle 0|,
\end{equation}
where $\rho_{AB}(0)$ denotes the initial state of the detectors, and $|0\rangle$ represents the vacuum state of the field. The dynamics of the total system are governed by the von Neumann equation,
\begin{equation}
\dot{\rho}_{\text{tot}}(\tau) = -i[H, \rho_{\text{tot}}(\tau)],
\end{equation}
where $\tau$ represents the proper time of the atoms. Given that the typical timescale of the environment is significantly shorter than that of the detectors, a Markovian evolution is assumed. Integrating out the background field's degrees of freedom yields the reduced dynamics of the detectors, driven by a quantum dynamical semigroup of completely positive maps. This evolution is represented by the Kossakowski-Lindblad master equation \cite{Gorini1976,Lindblad1976}:
\begin{equation}
\frac{\partial \rho_{AB}(\tau)}{\partial \tau} = -i[H_{\text{eff}}, \rho_{AB}(\tau)] + \mathcal{L}[\rho_{AB}(\tau)],
\end{equation}
where the dissipator $\mathcal{L}[\rho]$ accounts for the interaction with the external field:
\begin{equation}
\mathcal{L}[\rho] = \sum_{i,j=1,2,3} \sum_{\alpha, \beta=A,B} \frac{C_{ij}}{2} \left( 2\sigma_j^{(\beta)} \rho \sigma_i^{(\alpha)} - \{\sigma_i^{(\alpha)} \sigma_j^{(\beta)}, \rho \} \right).
\end{equation}
Here, $C_{ij}$ is the Kossakowski matrix. To define $C_{ij}$, we first introduce the Wightman function of the scalar field, which is given by:
\begin{equation}
G^+(x, x') = \langle 0 | \Phi(x) \Phi(x') | 0 \rangle.
\end{equation}
The Fourier transform of the Wightman function is expressed as:
\begin{equation}
G(\lambda) = \int_{-\infty}^{\infty} d\tau e^{i\lambda\tau} G^+(\tau) = \int_{-\infty}^{\infty} d\tau e^{i\lambda\tau} \langle \Phi(\tau) \Phi(0) \rangle.
\end{equation}
This Fourier transform allows for the determination of the coefficients $C_{ij}$ through the following decomposition:
\begin{equation}
C_{ij} = \frac{\gamma_+}{2}\delta_{ij} - i \frac{\gamma_-}{2}\epsilon_{ijk}\delta_{3,k} + \gamma_0\delta_{3,i}\delta_{3,j},
\end{equation}
where:
\begin{equation}\label{li1}
\gamma_\pm = G(\omega) \pm G(-\omega), \quad \gamma_0 = G(0) - \frac{\gamma_+}{2}.
\end{equation}
Moreover, the interaction with the external scalar field induces a Lamb shift contribution to the effective Hamiltonian:
\begin{equation}
H_{\text{eff}} = \frac{1}{2} \tilde{\omega} \sigma_3,
\end{equation}
where $\tilde{\omega} = \omega + i[K(-\omega) - K(\omega)]$, and $K(\lambda)$ is the Hilbert transform of the Wightman function:
\begin{equation}
K(\lambda) = \frac{1}{i\pi} P \int_{-\infty}^{\infty} d\omega \frac{G(\omega)}{\omega - \lambda}.
\end{equation}
Following the trajectory of the accelerating detectors, the field Wightman function adheres to the Kubo-Martin-Schwinger (KMS) condition:
\begin{equation}
G^+(\tau) = G^+(\tau + i\beta),
\end{equation}
where $\beta \equiv 1/T = 2\pi/a$. In the frequency domain, this relationship is expressed as:
\begin{equation}
G(\lambda) = e^{\beta \omega} G(-\lambda).
\end{equation}
Utilizing translation invariance $\langle 0 | \Phi(x(0)) \Phi(x(\tau)) | 0 \rangle = \langle 0 | \Phi(x(-\tau)) \Phi(x(0)) | 0 \rangle$ and subsequent algebraic manipulations, Eq.(\ref{li1}) determine:
\begin{equation}
\gamma_+ = \int_{-\infty}^{\infty} d\tau e^{i\lambda\tau} \langle 0 | \{\Phi(\tau), \Phi(0)\} | 0 \rangle = \left( 1 + e^{-\beta\omega} \right) G(\omega),
\end{equation}
\begin{equation}
\gamma_- = \int_{-\infty}^{\infty} d\tau e^{i\lambda\tau} \langle 0 | [\Phi(\tau), \Phi(0)] | 0 \rangle = \left( 1 - e^{-\beta\omega} \right) G(\omega).
\end{equation}
For subsequent analyses, we introduce the ratio:
\begin{equation}
\gamma \equiv \frac{\gamma_-}{\gamma_+} = \tanh \left( \frac{\beta\omega}{2} \right),
\end{equation}
which depends solely on the Unruh temperature $T$, and not on the local correlator of the background.

Considering the two-atom state in the Bloch representation, the reduced density matrix of the steady-state state of two Unruh-DeWitt (UDW) detectors at infinite-time limit can be derived, exhibiting an X-type structure \cite{Bhuvaneswari2022,Elghaayda2023}:
\begin{equation}
\rho_{AB} = \begin{pmatrix}
A & 0 & 0 & 0 \\
0 & C & D & 0 \\
0 & D & C & 0 \\
0 & 0 & 0 & B
\end{pmatrix},
\end{equation}
where:
\begin{equation}
A = \frac{(3 + \Delta_0)(\gamma - 1)^2}{4(3 + \gamma^2)}, \quad B = \frac{(3 + \Delta_0)(\gamma + 1)^2}{4(3 + \gamma^2)},
\end{equation}
\begin{equation}
C = \frac{3 - \Delta_0 - (\Delta_0 + 1)\gamma^2}{4(3 + \gamma^2)}, \quad D = \frac{\Delta_0 - \gamma^2}{2(3 + \gamma^2)}.
\end{equation}

It is observed that the final equilibrium state of the two-detector system is determined by the ratio $\gamma$, which reflects the thermal effects of the Unruh phenomenon, and the initial state parameters encapsulated by $\Delta_0 = \sum_i \text{Tr}[\rho_{AB}(0) \sigma_i^A \otimes \sigma_i^B]$. \(\Delta_0\) required to be within the range $-3 \leq \Delta_0 \leq 1$, guarantees the non-negativity of $\rho_{AB}(0)$.

\begin{figure*}[t]
\subfloat[$\omega=1$]{\includegraphics[width=0.32\linewidth]{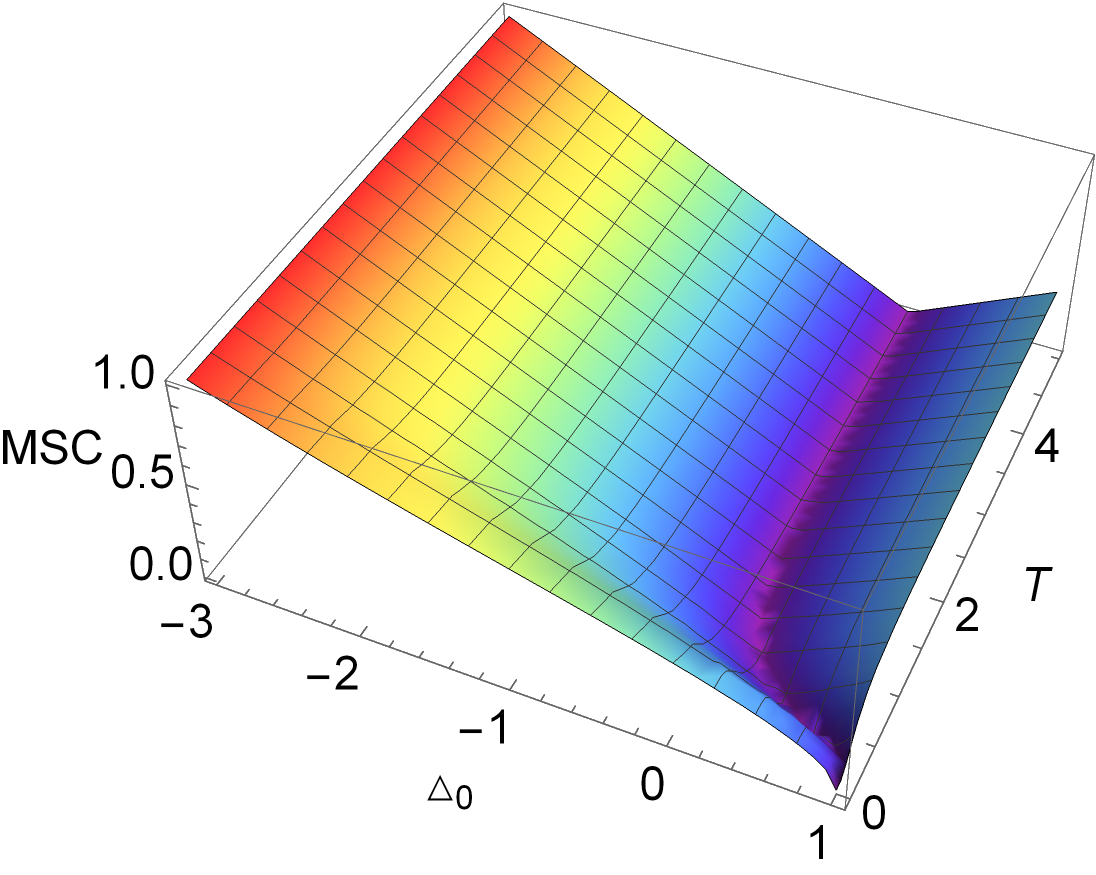}}
\subfloat[$\omega=3$]{\includegraphics[width=0.32\linewidth]{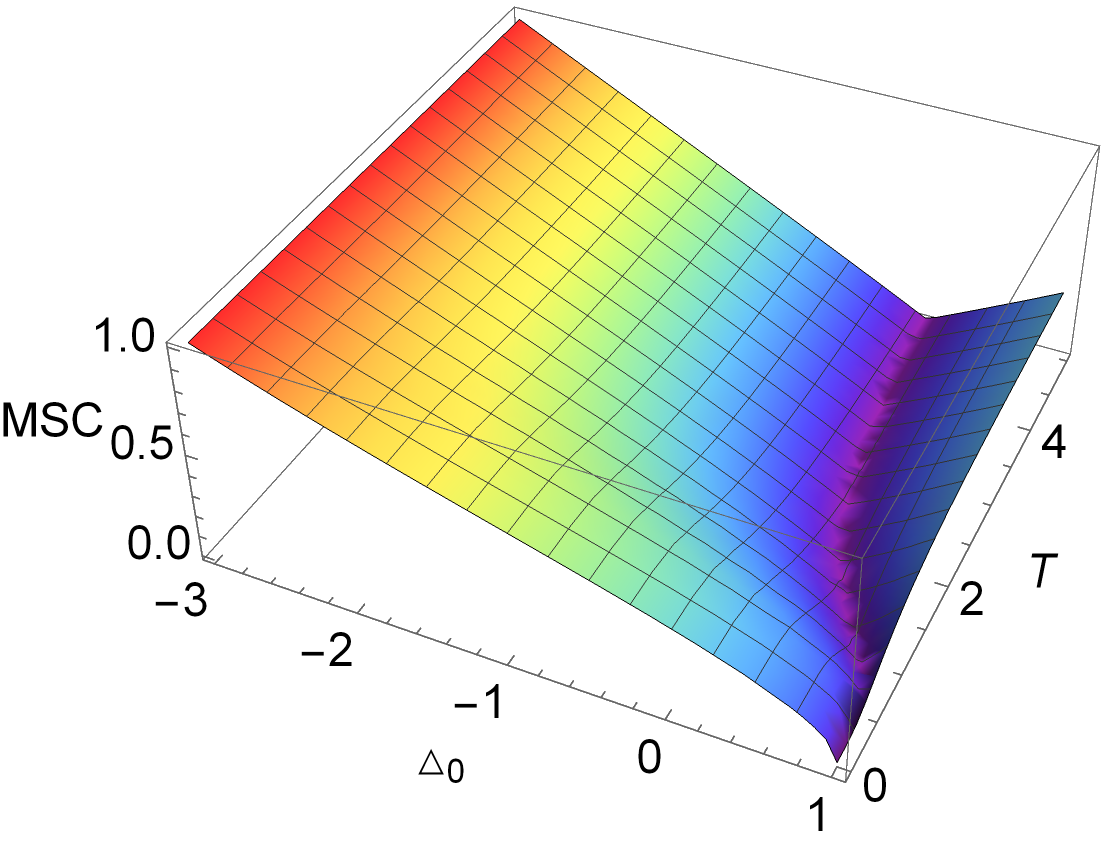}}
\subfloat[$\omega=5$]{\includegraphics[width=0.32\linewidth]{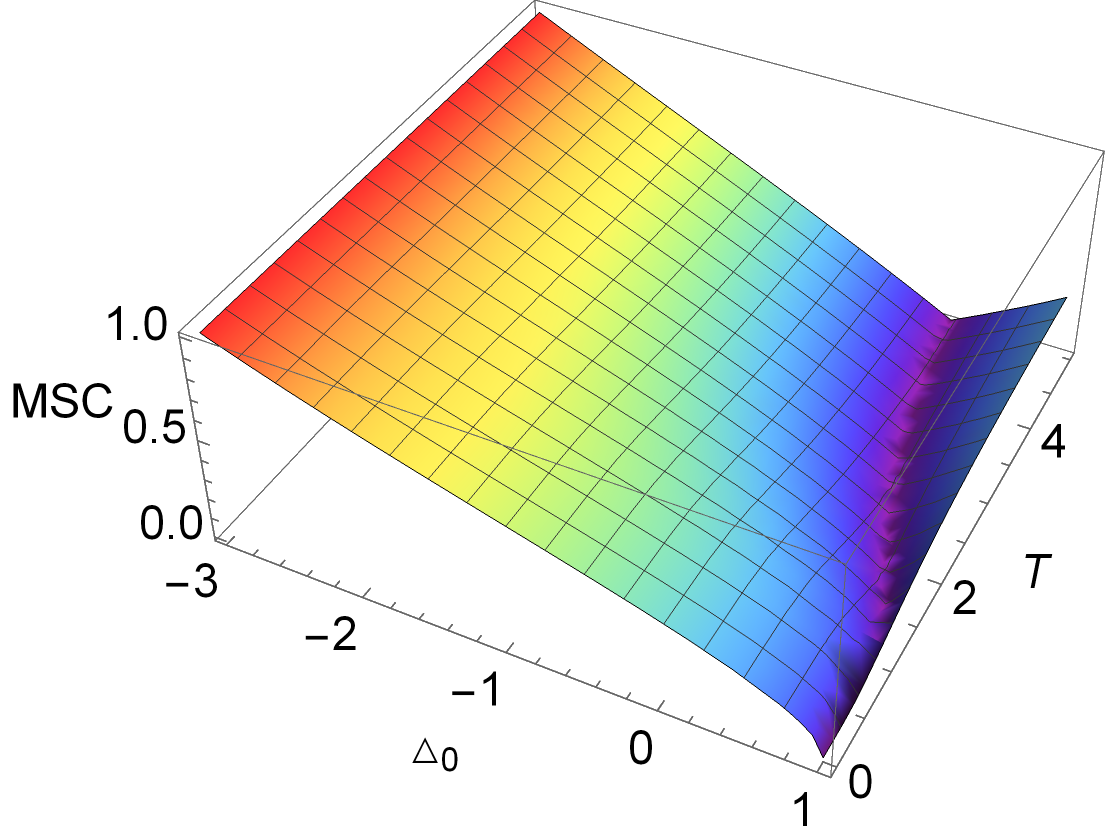}}
\caption{Maximal steered coherence of the Unruh-DeWitt detector as a function of the initial state selection parameter, $\Delta_0$, and the Unruh temperature, $T$, for different values of $\omega$: (a) $\omega=1$; (b) $\omega=3$; and (c) $\omega=5$.}
\label{fig1}
\end{figure*}
\begin{figure*}[t]
\subfloat[$\Delta_0=-1$]{\includegraphics[width=0.32\linewidth]{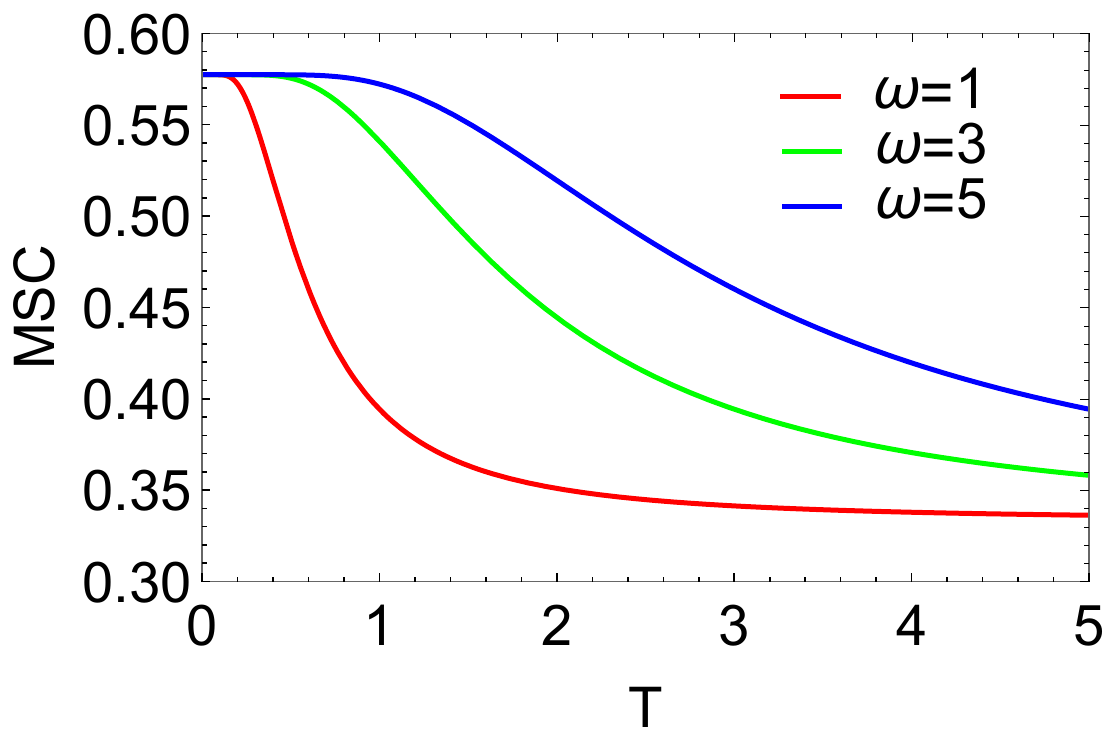}}
\subfloat[$\Delta_0=0.5$]{\includegraphics[width=0.32\linewidth]{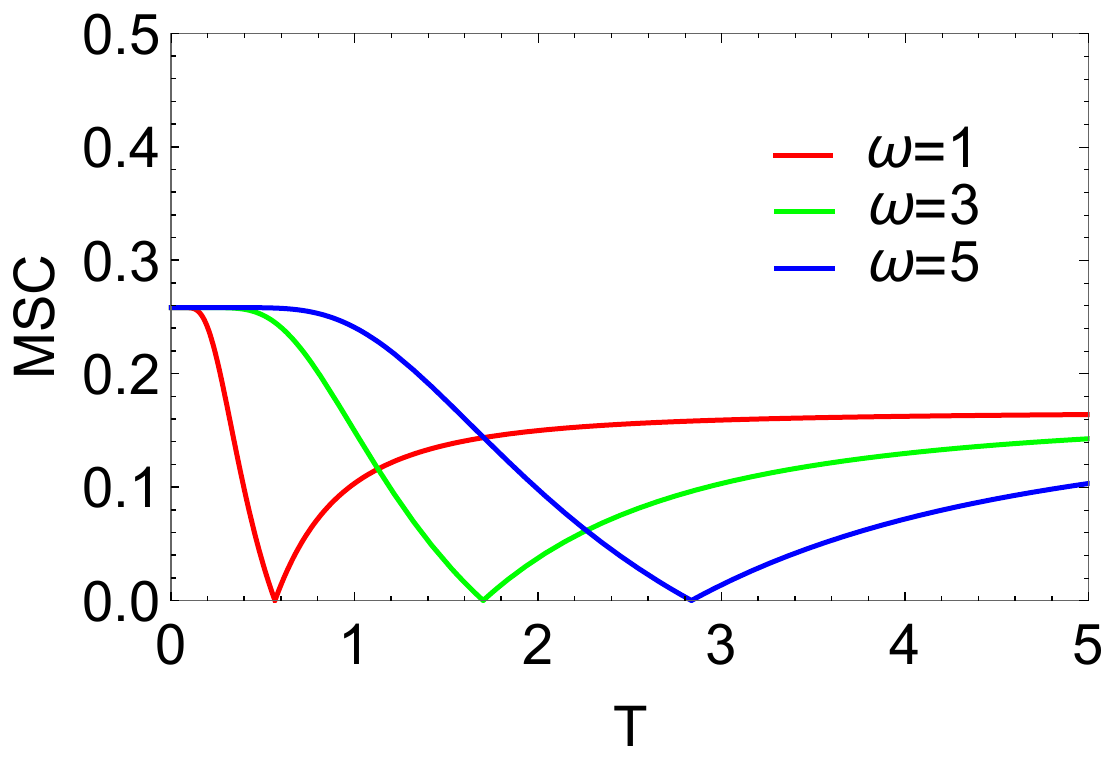}}
\subfloat[$\Delta_0=1$]{\includegraphics[width=0.32\linewidth]{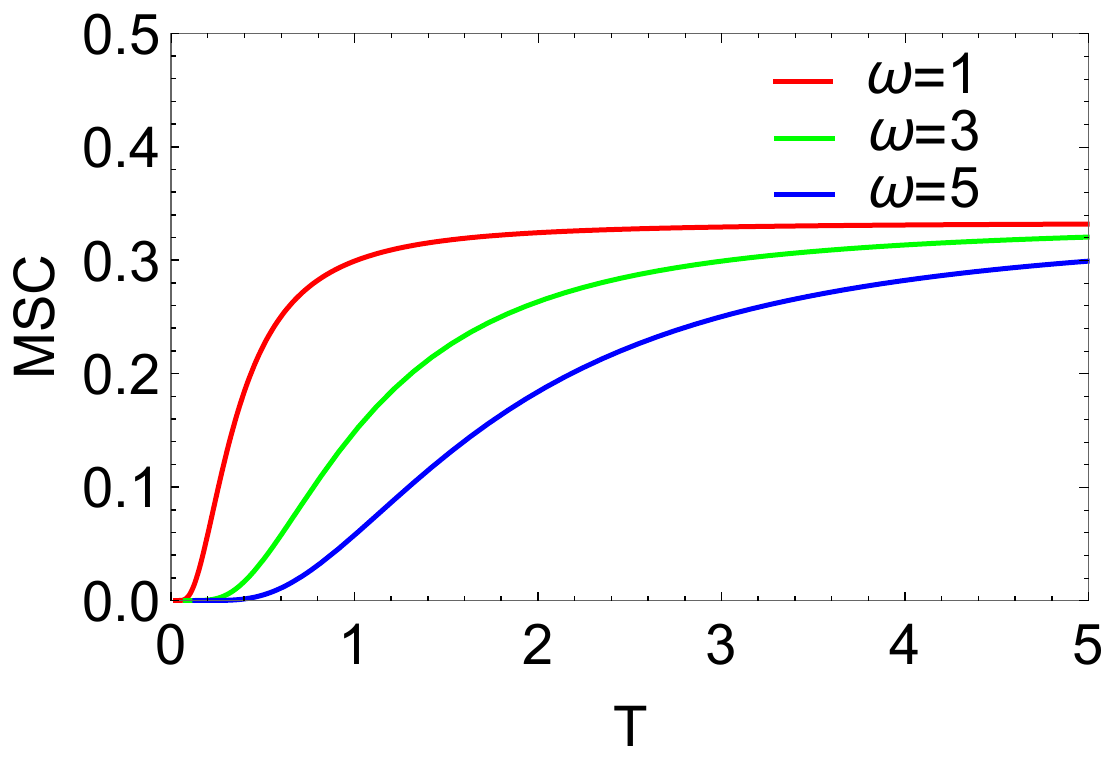}}
\caption{Maximal steered coherence of the Unruh-DeWitt detector as a function of the Unruh temperature $T$, for different values of the initial state selection parameter: (a)$\Delta_0 =-1$; (b)$\Delta_0 =0.5$; (c)$\Delta_0 =1$.}
\label{fig2}
\end{figure*}
\section{The maximal steered coherence in Unruh–deWitt detectors}\label{sec4}
In this section, we study the MSC fluctuations in quantum systems consisting of two accelerating UdW detectors associated with Minkowski space–time. Generally speaking, the Unruh effect is considered to be a kind of environmental decoherence, so we explore the impact of the Unruh temperature $T$, initial state choice parameter $\Delta_0$ and energy level spacing of the atom $\omega$ introduced in the previous section on the evolution of MSC.

To calculate MSC of $\rho_{AB}$, we needs to take the maximization over the set of projective measurements $M= (I+ \vec{m} \cdot \sigma)/2$, where $\sigma= (\sigma_{x}, \sigma_{y}, \sigma_{z})$ and $\vec{m} = (\sin \theta\cos \phi, \sin \theta\sin \phi, \cos \theta)$, with $\theta$ and $\phi$ being the polar and azimuth angles, respectively. Then, the post-measurement state of qubit $B$ can be obtained as

\begin{align}
\rho_{M_B}=\left(
             \begin{array}{cc}
               \frac{2(A+C+(A-C)\cos{\theta})}{p_{M}} & \frac{2De^{-i\phi}\sin{\theta}}{p_{M}} \\
               \frac{2De^{-i\phi}\sin{\theta}}{p_{M}} &  \frac{2(B+C+(C-B)\cos{\theta})}{p_{M}} \\
             \end{array}
           \right),
\end{align}
where $p_{M}=2(A+B+2C+(A-B)\cos(\theta))$. So the MSC can be obtained as a function of the initial state selection parameter $\Delta_0$ and the Unruh temperature $T$ for
\begin{align}\label{du1}
C_{max}(\rho_{AB})=\left|\frac{D}{\sqrt{(A+C)(C+B)}}\right|,
\end{align}
when $\theta=\mathrm{ArcCos}\left[\frac{B-A}{A+2C+B}\right]$ and $\phi$ can take any value.

In Fig. (\ref{fig1}), we present the MSC as a function of the initial state selection parameter, $\Delta_0$, and the Unruh temperature, $T$, for different fixed energy levels, $\omega$. From the figure, we observe distinct behaviors of MSC based on the values of $\Delta_0$:

\begin{itemize}
    \item For $-3 \leq \Delta_0 \leq 0$, the MSC decreases monotonically as the Unruh temperature $T$ increases.
    \item For $0 < \Delta_0 < 1$, the behavior of MSC depends on a threshold value for $T$:
    \begin{itemize}
        \item When $T < \frac{\omega}{2 \, \text{arctanh} \sqrt{\Delta_0}}$, the MSC decreases monotonically with increasing $T$.
        \item When $T > \frac{\omega}{2 \, \text{arctanh} \sqrt{\Delta_0}}$, the MSC increases monotonically with increasing $T$.
    \end{itemize}
    \item For $\Delta_0 = 1$, the MSC emerges as the temperature $T$ increases, eventually reaching a saturation value at higher temperatures.
\end{itemize}

We also find that as $T \rightarrow \infty$,
\begin{align}
    C_{max}(\rho_{AB})_{T \rightarrow \infty} = \frac{1}{3} |\Delta_0|.
\end{align}
This result indicates that, in the high-temperature limit, the MSC depends only on the initial state choice parameter, $\Delta_0$.

 In Fig. (\ref{fig2}), we present the MSC as a function of the Unruh temperature, \(T\), for different values of the initial state selection parameter, \(\Delta_0\), and fixed energy levels, \(\omega\). From Fig. \ref{fig2}(a), we observe that for \(\Delta_0 = -1\), MSC decreases steadily as the Unruh temperature \(T\) increases, indicating a monotonic decline. Specifically, as the energy level parameter \(\omega\) increases, the robustness of MSC against the rise in Unruh temperature is significantly enhanced, suggesting that higher energy levels help maintain coherence under thermal effects. In Fig. \ref{fig2}(b), for \(\Delta_0 = 0.5\), MSC exhibits a non-monotonic behavior: it initially decreases with increasing temperature \(T\), reaches a minimum, and then starts to rise. The temperature at which MSC begins to recover depends on the energy level \(\omega\). Higher energy levels shift the minimum point to higher temperatures and show a more pronounced recovery in MSC. In Fig. \ref{fig2}(c), for \(\Delta_0 = 1\), MSC increases with increasing temperature \(T\), suggesting that in this scenario, the Unruh effect actually aids in the stabilization of quantum coherence. Overall, the behavior of MSC under the influence of Unruh temperature varies significantly depending on the parameters involved. The Unruh effect plays a complex role in the evolution of quantum coherence—it may either suppress or enhance coherence depending on the initial state parameter \(\Delta_0\) and the system's energy level \(\omega\).

\section{Conclusion}\label{sec5}
In this work, we investigated the dynamics of two accelerating Unruh-DeWitt (UDW) detectors modeled as an open quantum system within the framework of a $3+1$-dimensional Minkowski spacetime. By treating the detectors as two-level systems interacting with a massless scalar field, we explored the effects of environmental decoherence and dissipation, focusing particularly on the role of the Unruh effect.

Our results reveal that the behavior of MSC is significantly influenced by the Unruh temperature $T$, energy level $\omega$, and the initial state parameter $\Delta_0$. For negative values, MSC generally decreases monotonically with an increase in the Unruh temperature, indicating that the Unruh effect acts as an environmental decoherence source. However, for $0<\Delta_0<1$, we observed non-monotonic behavior, where MSC initially decreases but subsequently recovers at higher temperatures, depending on the energy level  $\omega$. Interestingly, for $\Delta_0=1$, MSC can be generated.

These findings indicate that the Unruh effect has a nuanced impact on quantum coherence in open systems, with the ability to either suppress or enhance MSC depending on the system's initial state and energy parameters. Our study underscores the importance of careful selection of initial conditions and energy parameters to mitigate the decoherence effects of the Unruh phenomenon and potentially exploit its stabilizing influence on quantum coherence. This research contributes to a deeper understanding of coherence dynamics in relativistic quantum systems and may have implications for quantum technologies operating under extreme conditions.

\begin{acknowledgments}
This work was supported by the National Natural Science Foundation of China(Grant Nos. 12175106 and 92365110), the Natural Science Foundation of Jiangsu Province, China(Grant No.BK20240612), the Natural Science Research Start-up Foundation of Recruiting Talents of Nanjing University of Posts and Telecommunications(Grant No. NY222123), and the Natural Science Foundation of Nanjing University of Posts and Telecommunications(Grant No. NY223069).
\end{acknowledgments}

\end{document}